\documentclass[published]{JHEP3} 

\JHEP{00(2002)000}

\JHEPspecialurl{http://jhep.sissa.it/JOURNAL/JHEP3.tar.gz}

\usepackage{epsfig,multicol}

\newcommand\fverb{\setbox\pippobox=\hbox\bgroup\verb}
\newcommand\fverbdo{\egroup\medskip\noindent%
			\fbox{\unhbox\pippobox}\ }
\newcommand\fverbit{\egroup\item[\fbox{\unhbox\pippobox}]}
\newbox\pippobox

\def\sst{\scriptscriptstyle}

\def\d{\delta}

\def\e{\epsilon}
\def\z{\zeta}
\def\t{\theta}
\def\T{\Theta}

\def\o{\omega}

\def\be{\begin{equation}}
\def\ee{\end{equation}}
\def\ba{\begin{eqnarray}}
\def\ea{\end{eqnarray}}

\newcommand{\rpar}{\stackrel{\leftarrow}{\partial}}
\newcommand{\lpar}{\stackrel{\rightarrow}{\partial}}
\newcommand{\nn}{\nonumber\\}
\newcommand{\no}{\nonumber}

\title{Generalizations for Schouten-Nijenhuis bracket and for 
differential analog of special Yang-Baxter equations}

\author{Dmitrij V. Soroka and Vyacheslav A. Soroka\\
	Kharkov Institute of Physics and Technology, 61108 Kharkov, Ukraine\\
	E-mail: \email{dsoroka@kipt.kharkov.ua},
        \email{vsoroka@kipt.kharkov.ua}}
\received{} 		
\revised{}
\accepted{}		

\preprint{\hepth{9912999}}	

\abstract{The Schouten-Nijenhuis bracket is generalized for the superspace case
and for the Poisson brackets of opposite Grassmann parities. Quite a number of 
generalizations for the differential analog of the special Yang-Baxter 
equations is also proposed.}

\keywords{Superspaces, Differential and Algebraic Geometry, Integrable 
Equations in Physics}

\dedicated{Dedicated to the light memory of Anna Yakovlevna Gelukh (Kalaida)}

\begin{document} 


\section{Introduction}

Recently a prescription for the construction of new Poisson brackets from the 
bracket with a definite Grassmann parity was proposed \cite{ss}. This 
prescription is based on the use of exterior differentials of diverse 
Grassmann parities. It was indicated in~\cite{ss} that this prescription leads 
to the generalizations of the Schouten-Nijenhuis bracket~
\cite{scho,nij,nij1,fr-nij,kod-sp,but,bffls,oz} on the both superspace case 
and the case of the brackets with diverse Grassmann parities. In the present 
paper we give the details of these generalizations. Here we also develop 
quite a number of generalizations for the differential analog 
of the special\footnote{Concerning terminology see, for 
example,~\cite{karmas}.} Yang-Baxter equations.

\section{Poisson brackets related with the exterior differentials}

Let us recall the prescription for the construction from a given Poisson 
bracket of a Grassmann parity $\e\equiv0,1\pmod2$ of another one.

A Poisson bracket, having a Grassmann parity $\e$, written in arbitrary
non-canonical phase variables $z^a$
\ba\label{2.1}
\{A,B\}_\e=A\rpar_{z^a}\o_\e^{ab}(z)\lpar_{z^b}B,
\ea
where $\rpar$ and $\lpar$ are right and left derivatives respectively,
has the following main properties:
\ba
g(\{A,B\}_\e)\equiv g_A+g_B+\e\pmod2,\no
\ea
\ba
\{A,B\}_\e=-(-1)^{(g_A+\e)(g_B+\e)} \{B,A\}_\e,\no
\ea
\ba
\sum_{(ABC)}(-1)^{(g_A+\e)(g_C+\e)} \{A,\{B,C\}_\e\}_\e=0,\no
\ea
which lead to the corresponding relations for the matrix
$\o_\e^{ab}$
\be
g\left(\o_\e^{ab}\right)\equiv g_a+g_b+\e\pmod2,\label{2.2}
\ee
\be
\o_\e^{ab}=-(-1)^{(g_a+\e)(g_b+\e)} \o_\e^{ba},\label{2.3}
\ee
\be
\sum_{(abc)}(-1)^{(g_a+\e)(g_c+\e)} \o_\e^{ad}\partial_{z^d}
\o_\e^{bc}=0,\label{2.4}
\ee
where $\partial_{z^a}\equiv\partial/\partial z^a$ and $g_a\equiv g(z^a)$, 
$g_A\equiv g(A)$ are the corresponding Grassmann parities of phase coordinates
$z^a$ and a quantity $A$ and a sum with a symbol $(abc)$ under it
designates a summation over cyclic permutations of $a, b$ and $c$.
We shall consider the non-degenerated matrix $\o_\e^{ab}$ which has an inverse 
matrix $\o^\e_{ab}(-1)^{g_b\e}$ (a grading factor is chosen for the 
convenience)
\ba
\o_\e^{ab}\o^\e_{bc}(-1)^{g_c\e}=\d^a_c\no
\ea
(there is no summation over $\e$ in the previous relation) with the properties
\ba
g(\o^\e_{ab})\equiv g_a+g_b+\e\pmod2,\no
\ea
\ba
\o^\e_{ab}=(-1)^{(g_a+1)(g_b+1)}\o^\e_{ba},\no
\ea
\ba
\sum_{(abc)}(-1)^{(g_a+1)g_c}\partial_{z^a}\o^\e_{bc}=0.\no
\ea

The Hamilton equations for the phase variables $z^a$, which correspond
to a Hamiltonian $H_\e$ ($g(H_\e)=\e$),
\ba
{dz^a\over dt}=\{z^a,H_\e\}_\e=
\o_\e^{ab}\partial_{z^b}H_\e\label{2.5}
\ea
can be represented in the form
\ba
{dz^a\over dt}=\o_\e^{ab}\partial_{z^b}H_\e\equiv
\o_\e^{ab}{\partial(d_{\sst\z}H_\e)\over\partial(d_{\sst\z}z^b)}
\mathrel{\mathop=^{\rm def}}(z^a,d_{\sst\z}H_\e)_{\e+\z},\label{2.6}
\ea
where $d_{\sst\z}$ ($\z=0,1$) is one of the exterior differentials
$d_{\sst0}$ or $d_{\sst1}$, which have opposite Grassmann parities $0$ 
and $1$ respectively and following symmetry properties with respect to the 
ordinary multiplication
\ba\label{2.7}
d_{\sst 0}z^ad_{\sst 0}z^b=
(-1)^{g_ag_b}d_{\sst0}z^bd_{\sst0}z^a,
\ea
\ba\label{2.8}
d_{\sst1}z^ad_{\sst1}z^b=
(-1)^{(g_a+1)(g_b+1)}d_{\sst1}z^bd_{\sst1}z^a
\ea
and exterior products
\ba\label{2.9}
d_{\sst 0}z^a\wedge d_{\sst 0}z^b=
(-1)^{g_ag_b+1}d_{\sst0}z^b\wedge d_{\sst0}z^a,
\ea
\ba\label{2.10}
d_{\sst1}z^a\tilde\wedge d_{\sst1}z^b=
(-1)^{(g_a+1)(g_b+1)}d_{\sst1}z^b\tilde\wedge d_{\sst1}z^a.
\ea
We use different notations $\wedge$ and $\tilde\wedge$ for the exterior 
products of $d_{\sst 0}z^a$ and $d_{\sst1}z^a$ respectively. 

By taking the exterior differential
$d_{\sst\z}$ from the Hamilton equations (\ref{2.5}), we obtain
\ba
{d(d_{\sst\z}z^a)\over dt}=(d_{\sst\z}\o_\e^{ab})
{\partial(d_{\sst\z}H_\e)\over\partial(d_{\sst\z}z^b)}
+(-1)^{\z(g_a+\e)}\o_\e^{ab}\partial_{z^b}(d_{\sst\z}H_\e)
\mathrel{\mathop=^{\rm def}}(d_{\sst\z}z^a,d_{\sst\z}H_\e)_{\e+\z}.\label{2.11}
\ea
As a result of equations (\ref{2.6}) and (\ref{2.11}) we have by definition 
the following binary composition for functions $F$ and $H$ of the variables 
$z^a$ and their differentials $d_{\sst\z}z^a\equiv y_{\sst\z}^a$
\ba\label{2.12}
(F,H)_{\e+\z}=
F\Bigl[\rpar_{z^a}\o_\e^{ab}\lpar_{y_{\sst\z}^b}
&+(-1)^{\z(g_a+\e)}\rpar_{y_{\sst\z}^a}\o_\e^{ab}\lpar_{z^b}\cr\nn
&+\rpar_{y_{\sst\z}^a}y_{\sst\z}^c\left(\partial_{z^c}
\o_\e^{ab}\right)\lpar_{y_{\sst\z}^b}\Bigr]H.
\ea
By using relations (\ref{2.2})-(\ref{2.4}) for the matrix $\o_\e^{ab}$, we can 
establish the following properties for the binary composition (\ref{2.12})
\ba
g[(F,H)_{\e+\z}]\equiv g_F+g_H+\e+\z\pmod2,\no
\ea
\ba
(F,H)_{\e+\z}=-(-1)^{(g_F+\e+\z)(g_H+\e+\z)} (H,F)_{\e+\z},\no
\ea
\ba
\sum_{(EFH)}(-1)^{(g_E+\e+\z)(g_H+\e+\z)} (E,(F,H)_{\e+
\z})_{\e+\z}=0,\no
\ea
which mean that the composition (\ref{2.12}) satisfies all the main properties
for the Poisson bracket with the Grassmann parity equal to $\e+\z$. Thus,
the application of the exterior differentials of opposite Grassmann
parities to the given Poisson bracket results in the brackets of the
different Grassmann parities.

By transition to the co-differential variables $y^{\sst\e+\sst\z}_a$, related
with differentials $y_{\sst\z}^a$ by means of the matrix $\o_\e^{ab}$
\ba\label{2.13}
y_{\sst\z}^a=y^{\sst\e+\sst\z}_b\o_\e^{ba},
\ea
the Poisson bracket (\ref{2.12}) takes a canonical form\footnote{There is 
no summation over $\e$ in relation (\ref{2.13}).}
\ba\label{2.14}
(F,H)_{\e+\z}=F\left[\rpar_{z^a}\lpar_{y^{\sst\e+\sst\z}_a}
-(-1)^{g_a(g_a+\e+\z)}\rpar_{y^{\sst\e+\sst\z}_a}\lpar_{z^a}\right]H,
\ea
that can be proved with the use of the Jacobi identity (\ref{2.4}).

The bracket (\ref{2.12}) is given on the functions of the variables
$z^a$, $y_{\sst\z}^a$
\ba
F=\sum_p{1\over p!} y_{\sst\z}^{a_p}\cdots y_{\sst\z}^{a_1}
f_{a_1\ldots a_p}(z), \qquad 
g(f_{a_1\ldots a_p})=g_f+g_{a_1}+\cdots+g_{a_p},\label{2.15}
\ea
whereas this bracket, rewritten in the form (\ref{2.14}), is given on the 
functions of variables $z^a$ and $y^{\sst\e+\sst\z}_a$
\ba
F=\sum_p{1\over p!} y^{\sst\e+\sst\z}_{a_p}\cdots y^{\sst\e+\sst\z}_{a_1}
f^{a_1\ldots a_p}(z), \qquad 
g(f^{a_1\ldots a_p})=g_f+\e p+g_{a_1}+\cdots+g_{a_p}.\label{2.16}
\ea
We do not exclude a possibility of the own Grassmann parity $g_f\equiv g(f)$ 
for a quantity $f$. By taking into account relation (\ref{2.13}), we have the 
following rule for the rising of indices:
\ba
f^{b_1\ldots b_p}=(-1)^{\sum\limits_{k=1}^{p-1}
[g_{b_1}+\cdots+g_{b_k}+k(\e+\z)](g_{b_{k+1}}+g_{a_{k+1}}+\e)}
\o_\e^{b_pa_p}\cdots\o_\e^{b_1a_1}f_{a_1\ldots a_p}.\no
\ea
Note that the quantities $f_{a_1\ldots a_p}$ and $f^{a_1\ldots a_p}$ have 
in general the different symmetry and parity properties.

In the case $\z=1$, due to relations (\ref{2.8}), (\ref{2.10}), the terms in 
the decomposition of a function $F(z^a,y_{\sst1}^a)$ into degrees $p$ of the
variables $y_{\sst1}^a$ 
\ba
F=\sum_p{1\over p!} y_{\sst1}^{a_p}\cdots y_{\sst1}^{a_1}
f_{a_1\ldots a_p}(z)\label{2.17}
\ea
can be treated as $p$-forms and the bracket
(\ref{2.12}) can be considered as a Poisson bracket on $p$-forms so that
being taken between a $p$-form and a $q$-form results in a
$(p+q-1)$-form\footnote{Concerning a Poisson bracket between 1-forms and its 
relation with the Lie bracket of vector fields see in the book~\cite{stern}.}. 
Thus, the bracket (\ref{2.12}) is a generalization of the
bracket introduced in~\cite{karmas,kar} on the superspace case and on the 
case of the brackets (\ref{2.1}) with arbitrary Grassmann parities $\e$ 
($\e=0,1$).

\section{Generalizations of the Schouten-Nijenhuis bracket}

If we take the bracket in the canonical form (\ref{2.14}), then we obtain 
the generalizations of the Schouten-Nijenhuis bracket~\cite{scho,nij} 
(see also~\cite{nij1,fr-nij,kod-sp,but,bffls,oz,karmas}) onto the cases of 
superspace and the brackets of diverse Grassmann parities. Indeed, let us 
consider the bracket (\ref{2.14}) between monomials $F$ and $H$ having 
respectively degrees $p$ and $q$
\ba
F={1\over p!} y^{\sst\e+\sst\z}_{a_p}\cdots y^{\sst\e+\sst\z}_{a_1}
f^{a_1\ldots a_p}(z), \qquad 
g(f^{a_1\ldots a_p})=g_f+ p\e+g_{a_1}+\cdots+g_{a_p},\no
\ea
\ba
H={1\over q!} y^{\sst\e+\sst\z}_{a_q}\cdots y^{\sst\e+\sst\z}_{a_1}
h^{a_1\ldots a_q}(z), \qquad 
g(h^{a_1\ldots a_q})=g_h+q\e+g_{a_1}+\cdots+g_{a_q}.\no
\ea
Then as a result we obtain
\ba
(F,H)_{\e+\z}
&=&{(-1)^{[g_{b_1}+\cdots+g_{b_{q-1}}+(q-1)(\e+\z)](g_f+g_l+p\z)}
\over p!(q-1)!}\cr\nn
&\times& y^{\sst\e+\sst\z}_{b_{q-1}}\cdots y^{\sst\e+\sst\z}_{b_1}
y^{\sst\e+\sst\z}_{a_p}\cdots y^{\sst\e+\sst\z}_{a_1}
\left(f^{a_1\ldots a_p}\rpar_{z^l}\right)h^{b_1\ldots b_{q-1}l}\cr\nn
&-&{(-1)^{(g_l+\e+\z)(g_f+p\e+g_{a_2}+\cdots+g_{a_p})+
[g_{b_1}+\cdots+g_{b_q}+q(\e+\z)][g_f+\e+(p-1)\z]}\over (p-1)!q!}\cr\nn
&\times& y^{\sst\e+\sst\z}_{b_q}\cdots y^{\sst\e+\sst\z}_{b_1}
y^{\sst\e+\sst\z}_{a_p}\cdots y^{\sst\e+\sst\z}_{a_2}
f^{la_2\ldots a_p}\partial_{z^l}h^{b_1\ldots b_q}.\label{2.18}
\ea

\subsection{Particular cases}

Let us consider the formula (\ref{2.18}) for the particular values of $\e$
and $\z$. 

1. We start from the case which leads to the usual Schouten-Nijenhuis bracket 
for the skew-symmetric contravariant tensors. In this case, when $\e=0$, $\z=1$
and the matrix $\o_0^{ab}(x)=-\o_0^{ba}(x)$ corresponds to the usual Poisson
bracket for the commuting coordinates $z^a=x^a$, we have
\ba
(F,H)_1&=&{(-1)^{(q-1)(g_f+p)}\over p!(q-1)!}
\T_{b_{q-1}}\cdots\T_{b_1}\T_{a_p}\cdots\T_{a_1}
\left(f^{a_1\ldots a_p}\rpar_{x^l}\right)h^{b_1\ldots b_{q-1}l}\cr\nn
&-&{(-1)^{g_f(q+1)+q(p-1)}\over (p-1)!q!}
\T_{b_q}\cdots\T_{b_1}\T_{a_p}\cdots\T_{a_2}
f^{la_2\ldots a_p}\partial_{x^l}h^{b_1\ldots b_q},\label{3.1}
\ea
where $\T_a\equiv y_a^{\sst1}$ are Grassmann co-differential variables related 
owing to (\ref{2.13}) with the Grassmann differential variables 
$\T^a\equiv d_{\sst1}x^a$
\ba
\T^a=\T_b\o_0^{ba}.\no
\ea
When Grassmann parities of the quantities $f$ and $h$ are equal to zero
$g_f=g_h=0$, we obtain from (\ref{3.1})
\ba\label{3.2}
(F,H)_1\mathrel{\mathop=^{\rm def}}
(-1)^{(p+1)q+1}\T_{a_{p+q}}\cdots\T_{a_2}[F,H]^{a_2\ldots a_{p+q}},
\ea
where $[F,H]^{a_2\ldots a_{p+q}}$ are components of the usual 
Schouten-Nijenhuis bracket (see, for example, \cite{bffls}) for the 
contravariant antisymmetric tensors\footnote{Here and below we use the same 
notation [F,H] for the different brackets. We hope that this will not lead to 
the confusion.}. 
This bracket has the following symmetry property
\ba
[F,H]=(-1)^{pq}[H,F]\label{3.3}
\ea
and satisfies the Jacobi identity
\ba\label{3.4}
\sum_{(FHE)}(-1)^{ps}[[F,H],E]=0,
\ea
where $s$ is a degree of a monomial $E$.

2. In the case $\e=\z=0$ and $\o_0^{ab}(x)=-\o_0^{ba}(x)$ we obtain the 
bracket for symmetric contravariant tensors (see, for example, \cite{but})
\ba
(F,H)_0
&=&{1\over p!(q-1)!}
y^{\sst0}_{b_{q-1}}\cdots y^{\sst0}_{b_1}y^{\sst0}_{a_p}\cdots y^{\sst0}_{a_1}
\left(\partial_{x^l}f^{a_1\ldots a_p}\right)h^{b_1\ldots b_{q-1}l}\cr\nn
&-&{1\over(p-1)!q!}
y^{\sst0}_{b_q}\cdots y^{\sst0}_{b_1}y^{\sst0}_{a_p}\cdots y^{\sst0}_{a_2} 
f^{la_2\ldots a_p}\partial_{x^l}h^{b_1\ldots b_q}\mathrel{\mathop=^{\rm def}}
y^{\sst0}_{a_{p+q}}\cdots y^{\sst0}_{a_2}[F,H]^{a_2\ldots a_{p+q}},\cr\no
\ea
where commuting co-differentials $y^{\sst0}_a$ connected with commuting 
differentials $y^a_{\sst0}\equiv d_{\sst0}x^a$ in accordance with (\ref{2.13})
\ba
y_{\sst0}^a=y^{\sst0}_b\o_0^{ba}\no
\ea
and the bracket $[F,H]^{a_2\ldots a_{p+q}}$ has the following symmetry 
property
\ba
[F,H]=-(-1)^{g_fg_h}[H,F]\no
\ea
and satisfies the Jacobi identity
\ba
\sum_{(EFH)}(-1)^{g_eg_h}[E,[F,H]]=0.\no
\ea

3. By taking the Martin bracket \cite{mar} $\o_0^{ab}(\t)=\o_0^{ba}(\t)$ with 
Grassmann coordinates $z^a=\t^a$ $(g_a=1)$ as an initial bracket (\ref{2.1}), 
we have in the case $\z=0$ for antisymmetric contravariant tensors on the 
Grassmann algebra
\ba
(F,H)_0&=&{(-1)^{(q-1)(g_f+1)}\over p!(q-1)!}
\T_{b_{q-1}}\cdots\T_{b_1}\T_{a_p}\cdots\T_{a_1}
(f^{a_1\ldots a_p}\rpar_{\t^l})h^{b_1\ldots b_{q-1}l}\cr\nn
&+&{(-1)^{(q-1)g_f+p}\over (p-1)!q!}
\T_{b_q}\cdots\T_{b_1}\T_{a_p}\cdots\T_{a_2}
f^{la_2\ldots a_p}\partial_{\t^l}h^{b_1\ldots b_q}\cr\nn
&\stackrel{\rm def}{=}&
\T_{a_{p+q}}\cdots\T_{a_2}[F,H]^{a_2\ldots a_{p+q}},\cr\no
\ea
where the Grassmann co-differentials $\T_a$ related with the Grassmann 
differentials $\T^a$ as
\ba
d_{\sst0}\t^a\equiv\T^a=\T_b\o_0^{ba}.\no
\ea
The bracket $[F,H]$ has the following symmetry property
\ba
[F,H]=-(-1)^{g_fg_h}[H,F]\no
\ea
and satisfies the Jacobi identity
\ba
\sum_{(EFH)}(-1)^{g_eg_h}[E,[F,H]]=0.\no
\ea

4. By taking the Martin bracket again, in the case $\z=1$
\ba
d_{\sst1}\t^a\equiv x^a=x_b\o_0^{ba}\no
\ea
we obtain for the symmetric tensors on Grassmann algebra
\ba
(F,H)_1&=&{1\over p!(q-1)!}
x_{b_{q-1}}\cdots x_{b_1}x_{a_p}\cdots x_{a_1}
(f^{a_1\ldots a_p}\rpar_{\t^l})h^{b_1\ldots b_{q-1}l}\cr\nn
&-&{1\over (p-1)!q!}
x_{b_q}\cdots x_{b_1}x_{a_p}\cdots x_{a_2}
f^{la_2\ldots a_p}\partial_{\t^l}h^{b_1\ldots b_q}\cr\nn
&\stackrel{\rm def}{=}&
x_{a_{p+q}}\cdots x_{a_2}[F,H]^{a_2\ldots a_{p+q}}.\cr\no
\ea
The bracket $[F,H]$ has the following symmetry property
\ba
[F,H]=-(-1)^{(g_f+p+1)(g_h+q+1)}[H,F]\no
\ea
and satisfies the Jacobi identity
\ba
\sum_{(EFH)}(-1)^{(g_e+s+1)(g_h+q+1)}[E,[F,H]]=0.\no
\ea

5. In general, if we take the even bracket in superspace with coordinates 
$z^a=(x,\t)$, then in the case $\z=1$ we have
\ba
&(&F,H)_1={(-1)^{(g_{b_1}+\cdots+g_{b_{q-1}}+q-1)(g_f+g_l+p)}\over p!(q-1)!}
y^{\sst1}_{b_{q-1}}\cdots y^{\sst1}_{b_1}y^{\sst1}_{a_p}\cdots y^{\sst1}_{a_1}
(f^{a_1\ldots a_p}\rpar_{z^l})h^{b_1\ldots b_{q-1}l}\cr\nn
&-&{(-1)^{(g_l+1)(g_f+g_{a_2}+\cdots+g_{a_p})+
(g_{b_1}+\cdots+g_{b_q}+q)(g_f+p-1)}\over(p-1)!q!}
y^{\sst1}_{b_q}\cdots y^{\sst1}_{b_1}y^{\sst1}_{a_p}\cdots y^{\sst1}_{a_2}
f^{la_2\ldots a_p}\partial_{z^l}h^{b_1\ldots b_q}\cr\nn
&\stackrel{\rm def}{=}&
y^{\sst1}_{a_{p+q}}\cdots y^{\sst1}_{a_2}[F,H]^{a_2\ldots a_{p+q}},\cr\no
\ea
where
\ba
d_{\sst1}z^a\equiv y^a_{\sst1}=y^{\sst1}_b\o_0^{ba}.\no
\ea
The bracket $[F,H]$ has the following symmetry property
\ba
[F,H]=-(-1)^{(g_f+p+1)(g_h+q+1)}[H,F]\no
\ea
and satisfies the Jacobi identity
\ba
\sum_{(EFH)}(-1)^{(g_e+s+1)(g_h+q+1)}[E,[F,H]]=0.\no
\ea

6. In the case of the even bracket in superspace as initial one with $\z=0$ we 
obtain
\ba
(F,H)_0&=&{(-1)^{(g_{b_1}+\cdots+g_{b_{q-1}})(g_f+g_l)}\over p!(q-1)!}
y^{\sst0}_{b_{q-1}}\cdots y^{\sst0}_{b_1}y^{\sst0}_{a_p}\cdots y^{\sst0}_{a_1}
(f^{a_1\ldots a_p}\rpar_{z^l})h^{b_1\ldots b_{q-1}l}\cr\nn
&-&{(-1)^{g_l(g_f+g_{a_2}+\cdots+g_{a_p})+
g_f(g_{b_1}+\cdots+g_{b_q})}\over(p-1)!q!}
y^{\sst0}_{b_q}\cdots y^{\sst0}_{b_1}y^{\sst0}_{a_p}\cdots y^{\sst0}_{a_2}
f^{la_2\ldots a_p}\partial_{z^l}h^{b_1\ldots b_q}\cr\nn
&\stackrel{\rm def}{=}&
y^{\sst0}_{a_{p+q}}\cdots y^{\sst0}_{a_2}[F,H]^{a_2\ldots a_{p+q}},\cr\no
\ea
where
\ba
d_{\sst0}z^a\equiv y^a_{\sst0}=y^{\sst0}_b\o_0^{ba}.\no
\ea
The bracket $[F,H]$ has the following symmetry property
\ba
[F,H]=-(-1)^{g_fg_h}[H,F]\no
\ea
and satisfies the Jacobi identity
\ba
\sum_{(EFH)}(-1)^{g_eg_h}[E,[F,H]]=0.\no
\ea

7. Taking as an initial bracket the odd Poisson bracket in superspace with 
coordinates $z^a$, for the case $\z=0$ we have
\ba
(F,H)_1&=&{(-1)^{(g_{b_1}+\cdots+g_{b_{q-1}}+q-1)(g_f+g_l)}\over p!(q-1)!}
y^{\sst1}_{b_{q-1}}\cdots y^{\sst1}_{b_1}y^{\sst1}_{a_p}\cdots y^{\sst1}_{a_1}
(f^{a_1\ldots a_p}\rpar_{z^l})h^{b_1\ldots b_{q-1}l}\cr\nn
&-&{(-1)^{(g_l+1)(g_f+p+g_{a_2}+\cdots+g_{a_p})+
(g_f-1)(g_{b_1}+\cdots+g_{b_q}+q)}\over(p-1)!q!}\cr\nn&\times&
y^{\sst1}_{b_q}\cdots y^{\sst1}_{b_1}y^{\sst1}_{a_p}\cdots y^{\sst1}_{a_2}
f^{la_2\ldots a_p}\partial_{z^l}h^{b_1\ldots b_q}
\stackrel{\rm def}{=}
y^{\sst1}_{a_{p+q}}\cdots y^{\sst1}_{a_2}[F,H]^{a_2\ldots a_{p+q}},\cr\no
\ea
where
\ba
d_{\sst0}z^a\equiv y^a_{\sst0}=y^{\sst1}_b\o_1^{ba}.\no
\ea
The bracket $[F,H]$ has the following symmetry property
\ba
[F,H]=-(-1)^{(g_f+1)(g_h+1)}[H,F]\no
\ea
and satisfies the Jacobi identity
\ba
\sum_{(EFH)}(-1)^{(g_e+1)(g_h+1)}[E,[F,H]]=0.\no
\ea

8. At last for the odd Poisson bracket in superspace, taking as an initial 
one, we obtain in the case $\z=1$
\ba
(F,H)_0&=&(-1)^{(g_{b_1}+\cdots+g_{b_{q-1}})(g_f+p)}\biggl[{1\over p!(q-1)!}
y^{\sst0}_{b_{q-1}}\cdots y^{\sst0}_{b_1}y^{\sst0}_{a_p}\cdots y^{\sst0}_{a_1}
(f^{a_1\ldots a_p}\rpar_{z^l})h^{lb_1\ldots b_{q-1}}\cr\nn
&-&{(-1)^{(g_f+p)(g_l+g_{b_q})}\over(p-1)!q!}
y^{\sst0}_{b_q}\cdots y^{\sst0}_{b_1}y^{\sst0}_{a_p}\cdots y^{\sst0}_{a_2} 
f^{a_2\ldots a_pl}\partial_{z^l}h^{b_1\ldots b_q}\biggr]\cr\nn
&\stackrel{\rm def}{=}&
y^{\sst0}_{a_{p+q}}\cdots y^{\sst0}_{a_2}[F,H]^{a_2\ldots a_{p+q}},\cr\no
\ea
where
\ba
d_{\sst1}z^a\equiv y^a_{\sst1}=y^{\sst0}_b\o_1^{ba}.\no
\ea
The bracket $[F,H]$ has the following symmetry property
\ba
[F,H]=-(-1)^{(g_f+p)(g_h+q)}[H,F]\no
\ea
and satisfies the Jacobi identity
\ba
\sum_{(EFH)}(-1)^{(g_e+s)(g_h+q)}[E,[F,H]]=0.\no
\ea

Thus, we see that the formula (\ref{2.18}) contains as particular cases quite 
a number of the Schouten-Nijenhuis type brackets.

\section{Generalizations for the differential analog of the special 
Yang-Baxter equations}

The bracket (\ref{2.12}) for the monomials $F$ and $H$ having respectively 
degrees $p$ and $q$ and expressed in terms of the variables $y_{\sst\z}^a$ has 
the form
\ba
(F&,&H)_{\e+\z}
={(-1)^{[g_{b_1}+\cdots+g_{b_{q-1}}+(q-1)\z](q_f+\e+p\z+g_n)}
\over p!(q-1)!}\cr\nn&\times&
y_{\sst\z}^{b_{q-1}}\cdots y_{\sst\z}^{b_1}
y_{\sst\z}^{a_p}\cdots y_{\sst\z}^{a_1}
\left(f_{a_1\ldots a_p}\rpar_{z^l}\right)\o_\e^{ln}h_{b_1\ldots b_{q-1}n}\cr\nn
&+&{(-1)^{(g_{b_1}+\cdots+g_{b_q}+q\z)[g_f+\e+(p-1)\z]+\z(g_l+\e)+
(\z+g_l)(g_f+g_l+g_{a_2}+\cdots+g_{a_p})}\over(p-1)!q!}\cr\nn&\times&
y_{\sst\z}^{b_q}\cdots y_{\sst\z}^{b_1}y_{\sst\z}^{a_p}\cdots y_{\sst\z}^{a_2}
f_{la_2\ldots a_p}\o_\e^{ln}\partial_{z^n}h_{b_1\ldots b_q}\cr\nn
&+&{(-1)^{[g_{b_1}+\cdots+g_{b_{q-1}}+(q-1)\z](g_f+g_n+\e+p\z)+
(g_c+\z)[g_f+g_l+(p-1)\z]+(g_l+\z)(g_f+g_l+g_{a_2}+\cdots+g_{a_p})}
\over(p-1)!(q-1)!}\cr\nn&\times&
y_{\sst\z}^{b_{q-1}}\cdots y_{\sst\z}^{b_1}y_{\sst\z}^cy_{\sst\z}^{a_p}\cdots 
y_{\sst\z}^{a_2}f_{la_2\ldots a_p}\left(\partial_{z^l}\o_\e^{ln}\right)
h_{b_1\ldots b_{q-1}n}.\label{4.1}
\ea

The formulas (\ref{2.18}) and (\ref{4.1}) in particular case when $F=H\equiv S$
and the degree of the monomial $S$ is equal to two
\ba
S={1\over2}y^{\sst\e+\sst\z}_{a_2}y^{\sst\e+\sst\z}_{a_1}s^{a_1a_2}=
{1\over2}y_{\sst\z}^{a_2}y_{\sst\z}^{a_1}s_{a_1a_2}\no
\ea
take the following form
\ba
(S,S)_{\e+\z}&={1\over2}
(-1)^{(g_s+\e+\z)(g_l+g_{a_2}+g_{a_3})+g_s(\e+\z)}
\left[(-1)^{g_s}-(-1)^{\e+\z}\right]
y^{\sst\e+\sst\z}_{a_3}y^{\sst\e+\sst\z}_{a_2}y^{\sst\e+\sst\z}_{a_1}
s^{a_1l}\partial_{z^l}s^{a_2a_3}\cr\nn
&=(-1)^{(g_{a_2}+g_{a_3})(g_s+\z)+\z(g_s+g_l+\e)+g_l(g_s+1)}
y_{\sst\z}^{a_3}y_{\sst\z}^{a_2}y_{\sst\z}^{a_1}
\Bigl\{{(-1)^{(g_{a_2}+g_{a_3})\e}\left[(-1)^{g_s+\e+1}+(-1)^\z\right]\over2}
\cr\nn&\times 
s_{a_1l}\o_\e^{ln}\partial_{z^n}s_{a_2a_3}+(-1)^{g_{a_2}g_l+g_{a_3}\e+\z}
s_{a_1l}\left(\partial_{z^{a_2}}\o_\e^{ln}\right)s_{na_3}\Bigr\}.
\label{4.2}\ea

In order that the tensor $s^{ab}$ in (\ref{4.2}) itself should be a 
matrix for the Poisson bracket
\ba
(F,H)=F\rpar_{z^a}s^{ab}(z)\lpar_{z^b}H\no
\ea
and therefore satisfies the Jacobi identity, the bracket 
(\ref{4.2}) for $S$ has to be equal to zero
\ba\label{4.3}
(S,S)_{\e+\z}=0.
\ea

\subsection{Particular cases}

In two cases $g_s=0$, $\e+\z=0$ and $g_s=1$, $\e+\z=1$  relation (\ref{4.3})
is satisfied identically because of the symmetry property of the bracket
(\ref{4.2}).

1. When $g_s=0$, $\e=0$ and $\z=1$ we have the superspace generalization for 
the differential analog of the special Yang-Baxter equations
\ba
(S,S)_1
=(-1)^{g_{a_2}+g_{a_3}}
y_{\sst1}^{a_3}y_{\sst1}^{a_2}y_{\sst1}^{a_1}
\left[s_{a_1l}\o_0^{ln}\partial_{z^n}s_{a_2a_3}+(-1)^{g_lg_{a_2}}
s_{a_1l}\left(\partial_{z^{a_2}}\o_0^{ln}\right)s_{na_3}\right]=0.\no
\ea

2. In this case if $z^a=x^a$ are commuting coordinates and variables
$d_{\sst1}x^a=y_{\sst1}^a\equiv\t^a$ are Grassmann quantities, we obtain 
the well-known differential analog of the special Yang-Baxter equations (see, 
for example, \cite{karmas})
\ba
(S,S)_1=\t^{a_3}\t^{a_2}\t^{a_1}
\left[s_{a_1l}\o_0^{ln}\partial_{x^n}s_{a_2a_3}+
s_{a_1l}\left(\partial_{x^{a_2}}\o_0^{ln}\right)s_{na_3}\right]=0.\no
\ea

3. If $z^a=\t^a$ are Grassmann coordinates, variables 
$y_{\sst1}^a=d_{\sst1}\t^a\equiv x^a$ are commuting and the symmetric matrix
$\o_0^{ab}$ corresponds to the Martin bracket \cite{mar}, we have some 
generalization on Grassmann variables for the differential analog of the 
special Yang-Baxter equations
\ba
(S,S)_1=x^{a_3}x^{a_2}x^{a_1}
\left[s_{a_1l}\o_0^{ln}\partial_{\t^n}s_{a_2a_3}
-s_{a_1l}\left(\partial_{\t^{a_2}}\o_0^{ln}\right)s_{na_3}\right]=0.\no
\ea

4. In the case $g_s=0$, $\e=1$ and $\z=0$ we obtain
\ba
(S,S)_1&=&(-1)^{g_l}y_{\sst0}^{a_3}y_{\sst0}^{a_2}y_{\sst0}^{a_1}
\Bigl[(-1)^{g_{a_2}+g_{a_3}}s_{a_1l}\o_1^{ln}\partial_{z^n}s_{a_2a_3}\cr
&+&(-1)^{g_lg_{a_2}+g_{a_3}}s_{a_1l}\left(\partial_{z^{a_2}}\o_1^{ln}\right)
s_{na_3}\Bigr]=0\cr\no
\ea
that is another generalization of the above mentioned analog on the superspace 
case.

5. When $g_s=1$ and $\e=\z=1$ we have the third superspace generalization for 
this analog
\ba
(S,S)_0&=&-(-1)^{g_l}y_{\sst1}^{a_3}y_{\sst1}^{a_2}y_{\sst1}^{a_1}
\Bigl[(-1)^{g_{a_2}+g_{a_3}}s_{a_1l}\o_1^{ln}\partial_{z^n}s_{a_2a_3}\cr
&+&(-1)^{g_lg_{a_2}+g_{a_3}}s_{a_1l}\left(\partial_{z^{a_2}}\o_1^{ln}\right)
s_{na_3}\Bigr]=0.\cr\no
\ea

6. At last in the case when $g_s=1$ and $\e=\z=0$ we have the fourth 
superspace generalization
\ba
(S,S)_0=(-1)^{g_{a_2}+g_{a_3}}y_{\sst0}^{a_3}y_{\sst0}^{a_2}y_{\sst0}^{a_1}
\Bigl[s_{a_1l}\o_0^{ln}\partial_{z^n}s_{a_2a_3}
+(-1)^{g_lg_{a_2}}s_{a_1l}\left(\partial_{z^{a_2}}\o_0^{ln}\right)
s_{na_3}\Bigr]=0.\no
\ea
In the last case we have two sub-cases. 

6a) The first one when we take as an initial
bracket (\ref{2.1}) the usual Poisson bracket $\o_0^{ab}=-\o_0^{ba}$, 
$z^a\equiv x^a$ are commuting variables and 
$d_{\sst0}z^a=d_{\sst0}x^a\equiv y_{\sst0}^a$ are also commuting
\ba
(S,S)_0=y_{\sst0}^{a_3}y_{\sst0}^{a_2}y_{\sst0}^{a_1}
\Bigl[s_{a_1l}\o_0^{ln}\partial_{x^n}s_{a_2a_3}
+s_{a_1l}\left(\partial_{x^{a_2}}\o_0^{ln}\right)
s_{na_3}\Bigr]=0.\no
\ea

6b) In the second sub-case we take the Martin bracket on the Grassmann algebra 
as an initial one: $z^a\equiv\t^a$ are Grassmann variables, 
$\o_0^{ab}=\o_0^{ba}$ and $d_{\sst0}z^a=d_{\sst0}\t^a\equiv\T^a$ are also 
Grassmann quantities. In this sub-case we have
\ba
(S,S)_0=\T^{a_3}\T^{a_2}\T^{a_1}
\Bigl[s_{a_1l}\o_0^{ln}\partial_{\t^n}s_{a_2a_3}
-s_{a_1l}\left(\partial_{\t^{a_2}}\o_0^{ln}\right)
s_{na_3}\Bigr]=0.\no
\ea

Thus, we obtained quite a number of generalizations for the differential 
analog of the special Yang-Baxter equations.

\section{Conclusion}

We give the prescription for 
the construction from a given Poisson bracket of the definite Grassmann parity 
another bracket. For this construction we use the exterior differentials with 
different Grassmann parities. We proved  that the resulting Poisson bracket 
essentially depends on the parity of the exterior differential in spite of 
these differentials give the same exterior calculus \cite{ss}. The 
prescription leads to the set of different generalizations for the 
Schouten-Nijenhuis bracket. Thus, we see that the Schouten-Nijenhuis bracket 
and its possible generalizations are particular cases of the usual Poisson 
brackets of different Grassmann parities (\ref{2.14}). We hope that these 
generalizations will find their own application for the deformation 
quantization (see, for example, \cite{bffls,konts}) as well as the usual 
Schouten-Nijenhuis bracket.

We also proposed a lot of generalizations for the differential analog of the 
special Yang-Baxter equations. We also believe that these generalizations can 
be used for the description of integrable systems as well as the usual 
Yang-Baxter equations.

\section*{Acknowledgments}
We are sincerely grateful to J. Stasheff for the interest to the work and 
stimulating remarks. One of the authors (V.A.S.) sincerely thanks L. Bonora 
for the fruitful discussions and warm hospitality at the SISSA/ISAS (Trieste) 
where this work has been completed.


\begin{thebibliography}{999}
\bibitem{ss}D.V. Soroka and V.A. Soroka, \emph{Exterior differentials in 
superspace and Poisson brackets}, \jhep{0303}{2003}{001}; hep-th/0211280. 
\bibitem{scho}J.A. Schouten, \emph{Uber  differetialkomitanten zweier 
Kontravarianter  Grossen}, \newjournal{Proc.\ Nederl.\ Acad.\ Wetensh.,\
ser. A.\ }{pnawa}{43}{1940}{449}.
\bibitem{nij}A. Nijenhuis, \newjournal{Indag.\ Math.\ }{im}{17}{1955}{390}.
\bibitem{nij1}A. Nijenhuis, \newjournal{Proc.\ Kon.\ Ned.\ Akad.\ Wet.\
Amsterdam\ A.\ }{pknawaa}{58}{1955}{390}.
\bibitem{fr-nij}A. Frohlicher and A. Nijenhuis, \newjournal{Proc.\ Kon.\
Ned.\ Akad.\ Wet.\ Amsterdam\ A.\ }{pknawaa}{59}{1956}{338}.
\bibitem{kod-sp}K. Kodaira and D.C. Spencer, \am{74}{1961}{59}.
\bibitem{but} C. Buttin, \newjournal{Compt.\ Rend.\ Acad.\ Sci.\ Ser.\ A \ }
{crassa}{269}{1969}{87}.
\bibitem{bffls}F. Bayen, M. Flato, C. Fronsdal, A. Lichnerowicz,
D. Sternheimer, \emph{Deformation theory and quantization, 1. Deformations
of symplectic structures}, \ap{111}{1978}{61}.
\bibitem{oz}Z. Oziewicz, On Schouten-Nijenhuis and Frolicher-Nijenhuis
Lie modules, \emph{The lecture given at the XIX International Conference
on Differential Geometric Methods in Theoretical Physics}, Rapallo (Genova)
Italy, 1990.
\bibitem{karmas}M.V. Karasev and V.P. Maslov, \emph{Non-linear Poisson 
brackets. Geometry and quantization}, Moscow, Nauka, 1991.
\bibitem{stern}S. Sternberg, \emph{Lectures on differential geometry},
Prentice Hall, Inc. Englewood Cliffs, N.J. 1964.
\bibitem{kar}M.V. Karasev, \emph{Proceeding of the Conference
``Theory of group representations and its  applications in physics''},
Tambov, 1989; Moscow, Nauka, 1990.
\bibitem{mar} J.L. Martin, \emph{Generalized classical dynamics and the 
``classical analogue'' of a Fermi oscillator}, \newjournal{Proc.\ Roy.\ Soc.\ 
A \ }{prsa}{251}{1959}{536}.
\bibitem{konts}M. Kontsevich, \emph{Deformation quantization of Poisson 
manifolds, 1}, Preprint alg/9709040.


\end{thebibliography}
\end{document}